\documentclass{article}

\PassOptionsToPackage{numbers,sort&compress}{natbib}
\usepackage[preprint]{neurips_2026}

\usepackage[utf8]{inputenc}
\usepackage[T1]{fontenc}
\usepackage{hyperref}
\hypersetup{hidelinks}
\usepackage{url}
\usepackage{graphicx}
\usepackage{amsmath}
\usepackage{amssymb}
\usepackage{amsfonts}
\usepackage{algorithm}
\usepackage{algpseudocode}
\usepackage{booktabs}
\usepackage{multirow}
\usepackage{capt-of}
\usepackage{microtype}
\usepackage{xcolor}
\usepackage{enumitem}

\definecolor{accent}{HTML}{B3261E}
\definecolor{improve}{HTML}{1A56DB}

\newcommand{\tblscore}[2]{%
  \mbox{#1\,{\scriptsize$\pm$\,#2}}%
}

\newcommand{\tblbest}[2]{%
  \mbox{{\bfseries #1}\,{\bfseries\scriptsize$\boldsymbol{\pm}$\,#2}}%
}

\newcommand{\tblsecond}[2]{%
  \mbox{\underline{#1}\,{\scriptsize$\pm$\,#2}}%
}

\newlength{\blockdrop}

\AtBeginDocument{%
  }

\title{From Valid to Useful: Post-Verification Acquisition for Recursive Self-Improving Recommendation}

\author{%
  Tonmoy Hasan\thanks{Work done while working at Nubank.} \\
  Nubank \\
  {\footnotesize\texttt{thasan1@charlotte.edu}}
  \And
  Taylor Foust \\
  Nubank \\
  {\footnotesize\texttt{taylor.foust@nubank.com.br}}
  \And
  Shao Tang \\
  Nubank \\
  {\footnotesize\texttt{tang.shao@nubank.com.br}}
  \AND
  Leonardo Neves \\
  Nubank \\
  {\footnotesize\texttt{leonardo.neves@nubank.com.br}}
  \And
  Aman Gupta \\
  Nubank \\
  {\footnotesize\texttt{aman.gupta@nubank.com.br}}
  \And
  Hiroto Udagawa \\
  Nubank \\
  {\footnotesize\texttt{hiroto.udagawa@nubank.com.br}}
  \AND
  Helder Dias \\
  Nubank \\
  {\footnotesize\texttt{helder.dias@nubank.com.br}}
  \And
  Daniel Silva \\
  Nubank \\
  {\footnotesize\texttt{daniel.ribeiro@nubank.com.br}}
  \And
  Rohan Ramanath \\
  Nubank \\
  {\footnotesize\texttt{rohan.ramanath@nubank.com.br}}
}

\begin{document}

\maketitle

\begin{abstract}
Sequential recommenders can generate synthetic interaction sequences and retrain on the augmented corpus in a recursive self-improvement loop. To limit error accumulation, current methods verify each generated sequence remains predictive of the user's real interactions and discard those that drift away from it. Verification does not, however, determine which verified sequences should train the next model. With every verified sequence used for training, source sequences yielding more verified sequences or longer continuations have more influence, although neither quantity indicates how much those sequences will help the next model.

We formulate the decision of which verified sequences are used to train the next model as \emph{post-verification acquisition} and introduce {\bf Disagreement-Aware Recursive Self-Improving Recommendation (DA-RSIR)}. DA-RSIR caps each source sequence's contribution and ranks its verified sequences by how much the model's predictions disagree over their augmented interactions. It uses a score derived from Bayesian Active Learning by Disagreement (BALD) and estimated with Monte Carlo (MC) dropout. 

DA-RSIR requires no extra labels, teacher model, or quality scorer. Across four datasets, three recommender models, and two metrics, it improves on the retain-all approach in all $24$ comparisons and attains the highest mean in $23$ of $24$ overall; the aggregate improvement is statistically significant on both metrics. A single DA-RSIR round exceeds the retain-all approach's best gain over five recursive rounds. These findings establish post-verification acquisition as a separate control point in recursive self-improvement, separating which sequences pass verification from which verified sequences are used to train the next model.
\end{abstract}

\section{Introduction}
\label{sec:intro}

Sequential recommender systems predict a user's next interaction from their chronologically ordered history \citep{kang2018selfattentive}, but their effectiveness remains constrained by sparse and imbalanced interaction data \citep{idrissi2020systematic}. Augmentation and self-supervised methods add training data by masking, reordering, or otherwise perturbing observed interaction sequences \citep{dang2024survey,xie2022contrastive}, while generative methods synthesize interaction sequences that were never observed \citep{liu2023diffusion,yin2024dataset}. Recursive self-improvement goes one step further: the recommender generates synthetic sequences, a new model is trained on the enlarged corpus, and the process repeats without additional labels or an external teacher. Training a model on its own data can distort what it learns \citep{shumailov2024collapse,alemohammad2024self}. Keeping original data alongside synthetic data limits this effect \citep{gerstgrasser2024model}, but a model trained on its own predictions can still reinforce its mistakes \citep{arazo2020pseudo}.

Recursive Self-Improving Recommendation (RSIR) applies a fidelity check during generation \citep{zhang2026rsir}. A synthetic sequence grows from a source sequence one interaction at a time, and generation continues only while an unused later item from the source remains near the top of the model's predictions. In the first round, sources contain observed training-prefix interactions; at later rounds, they may also be synthetic sequences acquired earlier. This fidelity check therefore evaluates consistency with the current source sequence, not the downstream value of a candidate for successor model training. RSIR nevertheless retains every verified sequence, so source sequences yielding more candidates contribute more training examples, while longer continuations contribute more next-item prediction targets. Retaining every verified sequence is consequently an acquisition policy rather than a neutral consequence of verification. In our first-round experiments, RSIR yields higher mean Recall@10 than the same model trained only on real data in all twelve model--dataset settings, but lower mean NDCG@10 in four. This motivates treating candidate selection after verification as an explicit design choice.

We introduce \textbf{Dis\-agree\-ment-Aware Recursive Self-Improving Recommendation} (\textbf{DA-RSIR}), a post-verification acquisition framework that leaves RSIR's generation and verification unchanged. DA-RSIR limits each source sequence's contribution and ranks its verified sequences using disagreement estimated with MC dropout \citep{gal2016dropout}. Multiple forward passes with dropout measure how much the model's predictions disagree about the item at each generated position, corresponding to that item's contribution to the BALD score \citep{houlsby2011bald,gal2017deep}. DA-RSIR then samples representatives from the low-, middle-, and high-disagreement ranges. Acquisition requires no additional labels, external teacher, or separately trained quality model.

We evaluate DA-RSIR on four datasets and three recommender models under retraining and fine-tuning. In the main comparison, retraining with DA-RSIR achieves a higher mean than retain-all RSIR in all $24$ model--dataset--metric comparisons. Our contributions are threefold:
\begin{itemize}
\item We formulate post-verification acquisition as a distinct control problem in recursive self-improving recommendation. Verification checks a candidate's consistency with the source sequence; acquisition determines which verified candidates are used for next model training.

\item We introduce DA-RSIR, which preserves RSIR's generation and verification, ranks verified candidates using posterior disagreement over generated interactions, and samples candidates across disagreement ranges without an external teacher or separately trained quality model.

\item We evaluate DA-RSIR on four datasets and three recommender models under retraining and fine-tuning, achieving higher mean performance than retain-all RSIR in all $24$ model--dataset--metric comparisons under retraining.
\end{itemize}

\section{Related Work}
\label{sec:related}

\paragraph{\bf Sequential recommendation and data augmentation.}
Sequential recommendation has progressed through recurrent \citep{hidasi2016session} and convolutional \citep{tang2018personalized} encoders, unidirectional self-attention \citep{kang2018selfattentive}, bidirectional masked-item modeling \citep{sun2019bert4rec}, self-supervised pretraining \citep{zhou2020s3rec}, contrastive views of the same history \citep{xie2022contrastive}, frequency-enhanced hybrid attention \citep{du2023frequency}, and generative sequential transduction \citep{zhai2024actions}. Because performance is often limited by sparse interaction histories \citep{idrissi2020systematic}, data-augmentation methods construct additional training examples through sequence transformations \citep{xie2022contrastive,zhou2024contrastive} or learned generators \citep{liu2021augmenting,liu2023diffusion,yin2024dataset}. These methods determine how additional examples are generated; DA-RSIR instead operates after generation and verification, selecting which verified model-generated sequences become supervision for the successor model.

\paragraph{\bf Recursive self-improvement and self-consuming training.}
Learning from model-produced supervision spans semi-supervised classification \citep{sohn2020fixmatch,arazo2020pseudo} and iterative self-training in language models \citep{gulcehre2023rest}, but repeated use of model outputs can reinforce errors through confirmation bias \citep{arazo2020pseudo} or distort the learned distribution \citep{shumailov2024collapse,alemohammad2024self}. Methods therefore filter generated supervision before training: FixMatch retains only high-confidence pseudo-labels \citep{sohn2020fixmatch}, STaR uses rationales associated with correct answers \citep{zelikman2022star}, and Self-Instruct removes invalid and near-duplicate generations \citep{wang2023selfinstruct}. Training protocol also matters: retaining real data alongside synthetic data can prevent collapse in the settings studied \citep{gerstgrasser2024model}. RSIR applies fidelity verification to recommendation by stopping a continuation when no later item from its source sequence remains highly ranked \citep{zhang2026rsir}, yet retains every sequence that passes verification and basic filtering. Verification therefore determines which sequences pass the source-conditioned fidelity check, not which verified sequences should train the next model. DA-RSIR leaves verification unchanged and makes this selection explicit through post-verification acquisition.

\paragraph{\bf Posterior disagreement and data acquisition.}
Active learning selects informative examples from a pool for labeling, with BALD scoring them by mutual information between predictions and model parameters \citep{houlsby2011bald}. Monte Carlo dropout provides a tractable approximation in neural networks \citep{gal2016dropout,gal2017deep}. Batch acquisition must account for redundancy: selecting the highest-scoring examples independently can produce redundant batches that even underperform uniform acquisition \citep{sener2018active,kirsch2019batchbald}. This motivates avoiding a top-score-only acquisition rule rather than treating high disagreement as universally preferable. A parallel literature selects or values fixed-corpus examples using influence functions \citep{koh2017understanding}, Shapley values \citep{ghorbani2019datashapley}, target-task gradient similarity \citep{xia2024less}, or pruning criteria designed to improve scaling with retained dataset size \citep{sorscher2022beyond}. DA-RSIR instead selects among model-generated sequences that specify the next interaction at each generated position and have passed source-conditioned fidelity verification. Posterior disagreement complements this verification by guiding selection among these verified synthetic sequences.

\section{Preliminaries and Problem Formulation}
\label{sec:problem}

In this section, we formulate post-verification acquisition in recursive self-improvement. RSIR retains every sequence passing fidelity verification and basic filtering \citep{zhang2026rsir}; we treat retain-all as an acquisition policy.

\subsection{Sequential Recommendation}

Let $\mathcal{U}$ be the set of users and $\mathcal{I}$ the item catalog. For each user $u\in\mathcal{U}$, let $T^u$ denote the number of observed interactions and $s^u=(i_1^u,\ldots,i_{T^u}^u)$ the chronologically ordered interaction sequence, where $i_t^u\in\mathcal{I}$ is the item interacted with by user $u$ at position $t$. The initial real corpus is $\mathcal{D}_0=\{s^u:u\in\mathcal{U}\}$.

For any training sequence $s=(i_1,\ldots,i_{|s|})\in\mathcal{D}$, where $i_t\in\mathcal{I}$ is the item at position $t$, let $s_{<t}=(i_1,\ldots,i_{t-1})$ denote the preceding interactions. A sequential recommender $f_\theta$ defines a next-item distribution $p_\theta(i\mid s_{<t})$ over $i\in\mathcal{I}$. For a training corpus $\mathcal{D}$, the next-item prediction objective, optimized in practice using sampled softmax, is
\begin{equation}
\mathcal{L}_{\mathrm{NIP}}(\theta;\mathcal{D})
=
-\sum_{s\in\mathcal{D}}
\sum_{t=2}^{|s|}
\log p_\theta(i_t\mid s_{<t}).
\label{eq:nip}
\end{equation}

\subsection{Recursive Self-Improvement with Fidelity Control}

At recursive round $k$, the model $f_{\theta_k}$ is trained on the cumulative corpus
$\mathcal{D}_k=\mathcal{D}_0\cup\bigcup_{r=1}^{k}\mathcal{A}_r$,
where $\mathcal{A}_r$ is the synthetic increment acquired at round $r$. Thus, real sequences remain in the corpus while acquired synthetic sequences accumulate across rounds. Following RSIR \citep{zhang2026rsir}, every sequence in the current training corpus may serve as a source for generation, including synthetic sequences acquired in earlier rounds, and the generator makes up to a predefined number of attempts from each source.

For a source sequence $s$, candidate items at each generation step are drawn from the source with probability $p_{\mathrm{loc}}$ and from the global catalog $\mathcal{I}$ otherwise. Given the current context $c_t$, RSIR retains the $\kappa$ highest-scoring items, samples a provisional next item, and appends it to form $c'_t$. Let $\tau$ denote the fidelity threshold. The extension is accepted only if at least one unused source-sequence item remains within the top $\tau$ predictions:
\begin{equation}
\exists\, j\in s,\; j\notin c'_t
\quad\text{such that}\quad
\operatorname{Rank}_{\theta_k}(j\mid c'_t)\leq\tau.
\label{eq:fidelity}
\end{equation}
Thus, top-$\kappa$ sampling determines the proposed item, whereas fidelity verification determines whether the extension is accepted; the proposed item itself need not satisfy the top-$\tau$ condition. If accepted, generation continues from $c'_t$; otherwise, generation stops and the last verified context is retained. Candidates below the minimum required length are removed, and exact sequences are deduplicated.

Let $\mathcal{C}_{s,k+1}$ denote the verified candidates generated from source sequence $s$, and let
$\mathcal{C}^{\mathrm{fid}}_{k+1}=\bigcup_{s\in\mathcal{D}_k}\mathcal{C}_{s,k+1}$
denote the complete fidelity-valid bank. Each candidate retains its source-sequence identity and generation-attempt metadata for acquisition.

\subsection{Verification and Acquisition}

After verification forms the fidelity-valid bank $\mathcal{C}^{\mathrm{fid}}_{k+1}$, acquisition selects a subset
$\mathcal{A}_{k+1}\subseteq\mathcal{C}^{\mathrm{fid}}_{k+1}$
for successor training. The successor corpus is then
$\mathcal{D}_{k+1}=\mathcal{D}_k\cup\mathcal{A}_{k+1}$.
Because acquisition operates only on the verified bank, it cannot select a sequence rejected during verification. RSIR uses a retain-all policy, adding every sequence that survives fidelity verification and basic filtering.

\begin{figure*}[t]
    \centering
    \includegraphics[width=\textwidth]{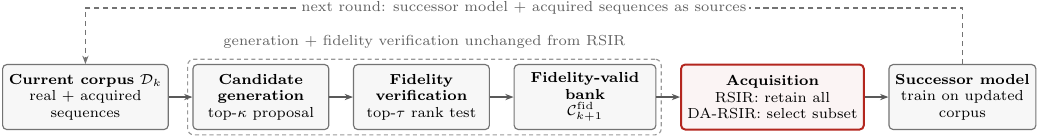}
    \caption{\textbf{Verification and acquisition in recursive self-improvement.} Candidate generation and fidelity verification produce a fidelity-valid candidate bank. Acquisition determines which verified candidates train the successor model: RSIR retains all, whereas DA-RSIR selects a subset. Acquired sequences enter the updated corpus and may serve as source sequences in the next recursive round.}
    \label{fig:verification_acquisition}
\end{figure*}

Under retain-all, every verified candidate is used for training, so each item after the first contributes a synthetic next-item prediction target. Source sequences that yield more verified candidates or longer continuations therefore contribute more training targets, although neither quantity indicates how much those candidates will help the successor model. Post-verification acquisition makes this contribution an explicit design decision. Because acquired sequences also enter the next round's training corpus, different selections from the same verified bank can affect both the successor model and future source sequences. Figure~\ref{fig:verification_acquisition} summarizes this separation between verification and acquisition within the recursive update.

\subsection{Post-Verification Acquisition Problem}
\label{sec:acquisition_problem}

Ideally, acquisition would select the verified candidates that produce the best successor model on real interactions. Let $\theta_{k+1}(\mathcal{A})$ denote the successor parameters obtained by training on the current corpus together with a selected subset $\mathcal{A}$. The ideal selection is
\begin{equation}
\mathcal{A}^{*}_{k+1}
\in
\operatorname*{arg\,min}_{
\mathcal{A}\subseteq
\mathcal{C}^{\mathrm{fid}}_{k+1}
}
\;
\mathbb{E}_{(x,i)\sim\mathcal{P}_{\mathrm{real}}}
\left[
-\log
p_{\theta_{k+1}(\mathcal{A})}(i\mid x)
\right],
\label{eq:ideal_acquisition}
\end{equation}
where $x$ is a real interaction context, $i$ its next item, and $\mathcal{P}_{\mathrm{real}}$ their unknown distribution. Optimizing this objective directly is impractical: the number of possible subsets grows exponentially with the size of the verified bank, and evaluating each subset requires training a successor model. We therefore seek a tractable acquisition policy; Section~\ref{sec:method} introduces DA-RSIR, which bounds each source sequence's contribution and organizes verified candidates using posterior disagreement.

\section{Disagreement-Aware Recursive Self-Improvement}
\label{sec:method}

DA-RSIR preserves RSIR's candidate generation and fidelity verification and changes only the subsequent acquisition step. For each source sequence, it limits how many verified candidates are selected, ranks them within proposal groups using posterior disagreement, and samples from low-, middle-, and high-disagreement ranges (Figure~\ref{fig:grouped_acquisition}).

\subsection{Target-Specific Posterior Disagreement}
\label{sec:disagreement}

Let $z=(i_1,\ldots,i_{|z|})$ denote a fidelity-verified candidate sequence. Following RSIR \citep{zhang2026rsir}, generation starts with the first source item, so items $i_t$ at positions $t\in\{2,\ldots,|z|\}$ are generated targets. Let $z_{<t}=(i_1,\ldots,i_{t-1})$ denote the preceding context.

We estimate posterior disagreement using Monte Carlo dropout \citep{gal2016dropout,gal2017deep}. During scoring, the current model is fixed, gradient computation is disabled, and dropout remains active. We perform $S$ stochastic forward passes, drawing a new dropout configuration for each pass and retaining it across all generated positions of the sequence so that each pass corresponds to a coherent approximate parameter draw. Let $p^{(r)}_{z,t}(j)$ denote the probability assigned to item $j$ at position $t$ in pass $r$. Before softmax, we mask the padding item and items already present in $z_{<t}$. The mean predictive probability is
$\overline{p}_{z,t}(j)=S^{-1}\sum_{r=1}^{S}p^{(r)}_{z,t}(j)$.

BALD measures the mutual information between predictions and model parameters at a given context \citep{houlsby2011bald}. It is large when the stochastic predictions are individually confident but disagree on the next item, and small when they agree or are uncertain in similar ways. Its Monte Carlo estimate can be written as
$\widehat{\mathrm{BALD}}_{z,t}=\sum_{j\in\mathcal{I}}b_{z,t}(j)$, where
\begin{equation}
b_{z,t}(j)
=
\frac{1}{S}\sum_{r=1}^{S}
p^{(r)}_{z,t}(j)\log p^{(r)}_{z,t}(j)
-
\overline{p}_{z,t}(j)\log\overline{p}_{z,t}(j).
\label{eq:target_contribution}
\end{equation}

For each generated position, DA-RSIR uses $b_{z,t}(i_t)$, the generated target's contribution to the MC BALD estimate. This target-specific term is not a mutual-information estimate on its own: it depends on both the probability assigned to $i_t$ and its variation across stochastic passes. Consequently, a sequence score formed from these terms need not produce the same candidate ordering as full BALD.

We average these contributions across the generated positions to obtain the sequence-level disagreement score
\begin{equation}
U_{\mathrm{mean}}(z)
=
\frac{1}{|z|-1}
\sum_{t=2}^{|z|}
b_{z,t}(i_t).
\label{eq:sequence_disagreement}
\end{equation}
Averaging prevents the score from growing automatically with the number of generated positions, although some sequence-length dependence may remain. For deterministic tie-breaking, we first use the maximum target-specific disagreement across generated positions and then the mean posterior-predictive log-likelihood of the generated targets. Any remaining ties are resolved using the generation-attempt index and a deterministic representation of the complete sequence. All quantities are computed only after fidelity verification, so they cannot affect candidate generation or verification decisions.

\subsection{Grouped Acquisition}
\label{sec:grouped_acquisition}

\begin{figure*}[t]
    \centering
    \includegraphics[width=0.95\textwidth]{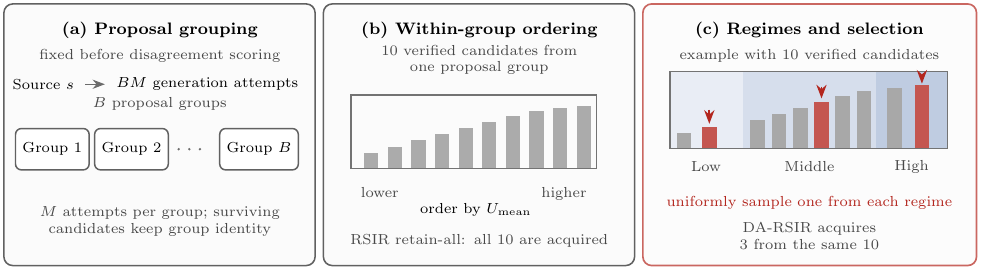}
    \caption{\textbf{Acquisition in DA-RSIR.}
    (a) Generation attempts from a source sequence are assigned to $B$ proposal groups before disagreement scoring.
    (b) Ten verified candidates from one group are shown in $U_{\mathrm{mean}}$ order; RSIR retains all ten.
    (c) DA-RSIR divides the same ten candidates into low-, middle-, and high-disagreement regimes and uniformly samples one representative from each.}
    \label{fig:grouped_acquisition}
\end{figure*}

\paragraph{\bf Proposal grouping.}
Generation attempts for source sequence $s$ are assigned by attempt index to $B$ disjoint proposal groups of $M$ attempts each, giving $BM$ attempts per source sequence. Grouping is fixed before disagreement scoring, so membership cannot depend on the realized scores, and each surviving candidate retains its group identity. This provides multiple proposal sets for the same source sequence while keeping comparisons within the same behavioral context. We treat $B$ as a design parameter selected using validation performance and include $B=1$ as the source-sequence-conditioned configuration without proposal grouping. Figure~\ref{fig:grouped_acquisition}(a) illustrates the assignment of generation attempts to proposal groups.

\paragraph{\bf Within-group ordering.}
Candidates in each nonempty group are ordered first by lower mean disagreement, then by lower maximum target-specific disagreement, and then by higher mean posterior-predictive log-likelihood. Remaining ties are resolved deterministically using the generation-attempt index and a deterministic representation of the complete sequence. Ranking separately within each source sequence and proposal group avoids requiring absolute disagreement magnitudes to be calibrated across heterogeneous source-sequence contexts. Figure~\ref{fig:grouped_acquisition}(b) shows ten verified candidates in disagreement order and highlights that RSIR retains all ten.

\paragraph{\bf Disagreement regimes.}
If a group has no verified candidates, it contributes none; if it has at most three, all are retained. For a group with $n>3$ candidates, let $h=\max\{1,\lfloor n/4\rfloor\}$. The first $h$ candidates form the low-disagreement regime, the last $h$ form the high-disagreement regime, and the remaining candidates form the middle-disagreement regime.

\paragraph{\bf Representative selection.}
For groups with more than three candidates, DA-RSIR samples one representative uniformly from each disagreement regime using a keyed pseudorandom seed determined by the experiment seed, recursive round, source sequence, proposal group, and regime. Uniform sampling preserves the intended regime without systematically selecting a rank boundary. The selected representatives across all source sequences and proposal groups form the synthetic increment after exact duplicate sequences are removed. Each group contributes at most three sequences, so each source sequence contributes at most $3B$, independently of how many verified candidates it produces. Retain-all instead keeps every verified candidate. DA-RSIR therefore limits how many sequences each source contributes, regardless of how many verified candidates it produces, while the number of next-item training targets still depends on selected-sequence lengths. Figure~\ref{fig:grouped_acquisition}(c) illustrates the disagreement regimes and representative selection from the same ten candidates shown in Figure~\ref{fig:grouped_acquisition}(b).

DA-RSIR samples across disagreement regimes rather than only the highest-scoring candidates. In batch active learning, independently selecting the highest-scoring BALD points can produce redundant batches and has sometimes underperformed uniform random acquisition \citep{kirsch2019batchbald}. This motivates avoiding a top-score-only rule but does not determine the ideal disagreement composition for recursive recommendation. Unlike BatchBALD, DA-RSIR does not maximize joint mutual information among selected candidates; its low, middle, and high regimes are relative positions in a source-sequence-conditioned ordering, not ordered levels of data quality. Thus, DA-RSIR does not assume that any regime is universally preferable. Algorithm~\ref{alg:darsir} in Appendix~\ref{app:algorithm} summarizes the acquisition procedure.

\subsection{Successor Training and Recursive Update}
\label{sec:successor_update}

After acquisition, the successor is trained on the updated corpus using its model-specific objective, whose next-item component is represented by Eq.~\eqref{eq:nip}. \textbf{DA-RSIR} retrains from a new random initialization, whereas \textbf{DA-RSIR-FT} initializes from the current model and fine-tunes on the same augmented corpus. We report the two procedures separately. Acquired sequences may serve as sources in the next round, so acquisition affects later generation through both the successor model and the next round's source corpus. Acquisition itself uses only the current model and the fidelity-verified bank, without training a successor model during selection.

\section{Experiments and Results}
\label{sec:experiments}

We evaluate post-verification acquisition across four datasets, three recommender models, two successor-training procedures, and single- and multi-round settings. Results are reported as four-run means and sample standard deviations. Appendix~\ref{app:experimental_details} provides the experimental protocol; Appendix~\ref{app:dataset_statistics} reports dataset and acquired-corpus statistics.

\paragraph{\bf Recommender models and baselines.} We evaluate all methods with three attention-based sequential-recommendation models: \textbf{SASRec}, based on unidirectional self-attention \citep{kang2018selfattentive}; \textbf{CL4SRec}, which combines sequential encoding with contrastive learning \citep{xie2022contrastive}; and \textbf{FEARec}, which incorporates frequency-domain information into attention \citep{du2023frequency}. \textbf{Base} denotes the corresponding model trained only on the real interaction corpus. We compare against \textbf{DA4Rec-SW} \citep{zhou2024contrastive}, \textbf{DiffuASR} \citep{liu2023diffusion}, \textbf{RSIR}, and \textbf{RSIR-FT} \citep{zhang2026rsir}, alongside \textbf{DA-RSIR} and \textbf{DA-RSIR-FT}.

\begin{table}[t]
\centering
\caption{First-round next-item recommendation results. Values are reported as mean $\pm$ sample standard deviation over four runs and are multiplied by $100$. N@10 and R@10 denote NDCG@10 and Recall@10, respectively. The highest mean is shown in bold and the second-highest distinct mean is underlined. Improv$_{\text{best}}$ reports the relative gain of the best DA-RSIR variant over the strongest baseline, and Improv$_{\text{base}}$ reports the relative gain of the same variant over Base. $^{*}$ indicates statistical significance at $p<0.05$; $^{\circ}$ indicates a non-significant comparison.}
\label{tab:main-fourseed}

\begingroup
\scriptsize
\renewcommand{\arraystretch}{1.03}
\setlength{\tabcolsep}{1.35pt}

\begin{tabular*}{\textwidth}{@{\extracolsep{\fill}}ll*{8}{c}@{}}
\toprule
& & \multicolumn{2}{c}{Amazon Toys} & \multicolumn{2}{c}{Amazon Beauty} & \multicolumn{2}{c}{Amazon Sports} & \multicolumn{2}{c}{Yelp} \\
\cmidrule(lr){3-4}
\cmidrule(lr){5-6}
\cmidrule(lr){7-8}
\cmidrule(lr){9-10}
Model & Method & N@10 & R@10 & N@10 & R@10 & N@10 & R@10 & N@10 & R@10 \\
\midrule

\multirow{9}{*}{\textbf{SASRec}}
& Base
& \tblscore{5.67}{0.09} & \tblscore{9.30}{0.12}
& \tblsecond{5.08}{0.07} & \tblscore{8.59}{0.04}
& \tblscore{2.73}{0.01} & \tblscore{4.69}{0.02}
& \tblscore{1.96}{0.07} & \tblscore{3.84}{0.09} \\
& DA4Rec-SW
& \tblscore{5.50}{0.03} & \tblscore{9.02}{0.13}
& \tblscore{4.92}{0.03} & \tblscore{8.27}{0.11}
& \tblscore{2.68}{0.03} & \tblscore{4.61}{0.08}
& \tblscore{1.81}{0.07} & \tblscore{3.60}{0.17} \\
& DiffuASR
& \tblscore{5.68}{0.07} & \tblscore{9.28}{0.09}
& \tblscore{5.00}{0.09} & \tblscore{8.45}{0.14}
& \tblscore{2.74}{0.06} & \tblscore{4.70}{0.13}
& \tblscore{1.96}{0.01} & \tblscore{3.85}{0.03} \\
& RSIR
& \tblscore{5.68}{0.06} & \tblscore{9.56}{0.07}
& \tblscore{5.01}{0.07} & \tblscore{8.67}{0.15}
& \tblscore{2.85}{0.01} & \tblscore{5.02}{0.06}
& \tblscore{2.17}{0.04} & \tblscore{4.22}{0.06} \\
& RSIR-FT
& \tblsecond{5.73}{0.07} & \tblsecond{9.64}{0.13}
& \tblscore{5.04}{0.08} & \tblscore{8.70}{0.06}
& \tblscore{2.86}{0.03} & \tblsecond{5.04}{0.06}
& \tblsecond{2.18}{0.06} & \tblscore{4.26}{0.07} \\
& DA-RSIR
& \tblscore{5.70}{0.02} & \tblscore{9.59}{0.06}
& \tblbest{5.17}{0.02} & \tblbest{8.86}{0.13}
& \tblbest{2.90}{0.03} & \tblbest{5.08}{0.04}
& \tblbest{2.21}{0.06} & \tblsecond{4.29}{0.05} \\
& DA-RSIR-FT
& \tblbest{5.77}{0.04} & \tblbest{9.67}{0.06}
& \tblbest{5.17}{0.06} & \tblsecond{8.81}{0.05}
& \tblsecond{2.88}{0.02} & \tblsecond{5.04}{0.06}
& \tblbest{2.21}{0.07} & \tblbest{4.32}{0.10} \\
\cmidrule(lr){2-10}
& Improv$_{\text{best}}$
& +0.7\%$^{*}$ & +0.3\%$^{*}$
& +1.8\%$^{*}$ & +1.8\%$^{*}$
& +1.4\%$^{*}$ & +0.8\%$^{*}$
& +1.4\%$^{*}$ & +1.4\%$^{*}$ \\
& Improv$_{\text{base}}$
& +1.8\%$^{*}$ & +4.0\%$^{*}$
& +1.8\%$^{*}$ & +3.1\%$^{*}$
& +6.2\%$^{*}$ & +8.3\%$^{*}$
& +12.8\%$^{*}$ & +12.5\%$^{*}$ \\

\midrule

\multirow{9}{*}{\textbf{CL4SRec}}
& Base
& \tblscore{5.50}{0.04} & \tblscore{9.00}{0.08}
& \tblscore{4.89}{0.05} & \tblscore{8.32}{0.05}
& \tblscore{2.60}{0.05} & \tblscore{4.50}{0.09}
& \tblscore{1.96}{0.06} & \tblscore{3.84}{0.05} \\
& DA4Rec-SW
& \tblscore{5.50}{0.08} & \tblscore{9.06}{0.11}
& \tblscore{4.84}{0.08} & \tblscore{8.28}{0.09}
& \tblscore{2.54}{0.03} & \tblscore{4.50}{0.07}
& \tblscore{1.97}{0.05} & \tblscore{3.83}{0.10} \\
& DiffuASR
& \tblbest{5.67}{0.10} & \tblscore{9.22}{0.08}
& \tblscore{4.84}{0.02} & \tblscore{8.26}{0.05}
& \tblscore{2.50}{0.07} & \tblscore{4.35}{0.09}
& \tblscore{1.95}{0.02} & \tblscore{3.79}{0.05} \\
& RSIR
& \tblscore{5.45}{0.06} & \tblscore{9.22}{0.06}
& \tblscore{4.87}{0.10} & \tblsecond{8.55}{0.12}
& \tblsecond{2.79}{0.05} & \tblsecond{4.90}{0.08}
& \tblscore{2.04}{0.02} & \tblsecond{4.00}{0.07} \\
& RSIR-FT
& \tblscore{5.55}{0.03} & \tblscore{9.19}{0.09}
& \tblscore{4.84}{0.02} & \tblscore{8.45}{0.11}
& \tblscore{2.75}{0.06} & \tblscore{4.84}{0.09}
& \tblscore{2.04}{0.03} & \tblscore{3.96}{0.01} \\
& DA-RSIR
& \tblscore{5.49}{0.05} & \tblbest{9.30}{0.15}
& \tblbest{4.96}{0.08} & \tblbest{8.64}{0.10}
& \tblbest{2.84}{0.04} & \tblbest{5.01}{0.10}
& \tblbest{2.09}{0.03} & \tblbest{4.05}{0.07} \\
& DA-RSIR-FT
& \tblsecond{5.57}{0.03} & \tblsecond{9.23}{0.03}
& \tblsecond{4.91}{0.09} & \tblscore{8.53}{0.09}
& \tblscore{2.71}{0.06} & \tblscore{4.81}{0.10}
& \tblsecond{2.06}{0.05} & \tblsecond{4.00}{0.08} \\
\cmidrule(lr){2-10}
& Improv$_{\text{best}}$
& $-$1.8\%$^{\circ}$ & +0.9\%$^{*}$
& +1.4\% $^{*}$& +1.1\%$^{*}$
& +1.8\% $^{*}$& +2.2\%$^{*}$
& +2.5\% $^{*}$& +1.3\%$^{*}$ \\
& Improv$_{\text{base}}$
& +1.3\%$^{*}$ & +3.3\%$^{*}$
& +1.4\%$^{*}$ & +3.8\%$^{*}$
& +9.2\%$^{*}$ & +11.3\%$^{*}$
& +6.6\%$^{*}$ & +5.5\%$^{*}$ \\

\midrule

\multirow{9}{*}{\textbf{FEARec}}
& Base
& \tblscore{6.08}{0.12} & \tblscore{9.67}{0.18}
& \tblscore{5.62}{0.09} & \tblscore{9.19}{0.15}
& \tblscore{3.13}{0.13} & \tblscore{5.28}{0.24}
& \tblscore{1.89}{0.06} & \tblscore{3.78}{0.13} \\
& DA4Rec-SW
& \tblscore{6.03}{0.12} & \tblscore{9.57}{0.22}
& \tblsecond{5.66}{0.07} & \tblsecond{9.37}{0.09}
& \tblscore{3.21}{0.05} & \tblscore{5.48}{0.08}
& \tblscore{1.86}{0.06} & \tblscore{3.69}{0.13} \\
& DiffuASR
& \tblscore{6.04}{0.15} & \tblscore{9.55}{0.22}
& \tblscore{5.62}{0.08} & \tblscore{9.20}{0.12}
& \tblscore{3.04}{0.10} & \tblscore{5.16}{0.14}
& \tblscore{1.91}{0.05} & \tblscore{3.78}{0.11} \\
& RSIR
& \tblsecond{6.25}{0.10} & \tblsecond{10.20}{0.18}
& \tblscore{5.56}{0.05} & \tblscore{9.28}{0.09}
& \tblsecond{3.26}{0.09} & \tblsecond{5.59}{0.09}
& \tblsecond{2.06}{0.01} & \tblsecond{4.12}{0.06} \\
& RSIR-FT
& \tblscore{6.11}{0.14} & \tblscore{9.85}{0.21}
& \tblscore{5.51}{0.14} & \tblscore{9.17}{0.25}
& \tblscore{3.21}{0.11} & \tblscore{5.51}{0.21}
& \tblscore{2.00}{0.04} & \tblscore{3.98}{0.08} \\
& DA-RSIR
& \tblbest{6.34}{0.07} & \tblbest{10.33}{0.06}
& \tblbest{5.67}{0.02} & \tblbest{9.52}{0.08}
& \tblbest{3.31}{0.06} & \tblbest{5.68}{0.10}
& \tblbest{2.12}{0.01} & \tblbest{4.19}{0.03} \\
& DA-RSIR-FT
& \tblscore{6.17}{0.21} & \tblscore{9.96}{0.33}
& \tblscore{5.54}{0.19} & \tblscore{9.26}{0.32}
& \tblscore{3.21}{0.19} & \tblscore{5.50}{0.35}
& \tblscore{2.01}{0.07} & \tblscore{4.05}{0.15} \\
\cmidrule(lr){2-10}
& Improv$_{\text{best}}$
& +1.4\%$^{*}$ & +1.3\%$^{*}$
& +0.2\%$^{*}$ & +1.6\%$^{*}$
& +1.5\%$^{*}$ & +1.6\%$^{*}$
& +2.9\%$^{*}$ & +1.7\%$^{*}$ \\
& Improv$_{\text{base}}$
& +4.3\%$^{*}$ & +6.8\%$^{*}$
& +0.9\%$^{*}$ & +3.6\%$^{*}$
& +5.8\%$^{*}$ & +7.6\%$^{*}$
& +12.2\%$^{*}$ & +10.8\%$^{*}$ \\

\bottomrule
\end{tabular*}
\endgroup
\end{table}

\subsection{Overall Performance}
\label{sec:main_results}

Table~\ref{tab:main-fourseed} reports performance after the first recursive self-improvement round. A DA-RSIR variant attains the highest mean in $23$ of the $24$ model--dataset--metric columns, covering every Recall@10 column and all but one NDCG@10 column. Relative to Base, DA-RSIR under retraining improves NDCG@10 by $5.1\%$ and Recall@10 by $6.6\%$ on average.

Compared with RSIR, DA-RSIR performs better in all $24$ model--dataset--metric comparisons under retraining, and all improvements are statistically significant at $p<0.05$. DA-RSIR-FT also improves over RSIR-FT in most fine-tuning settings and remains competitive in the others. Each matched pair uses the same candidate-generation and fidelity-verification configuration, differing only in how the fidelity-valid candidates are acquired for successor training. Retraining is generally stronger for CL4SRec and FEARec, whereas fine-tuning is more competitive for SASRec, motivating evaluation under both procedures.

The sole exception to the overall pattern is NDCG@10 for CL4SRec on Amazon Toys, where DiffuASR has the highest mean and DA-RSIR-FT ranks second. The difference is not statistically resolved (paired two-sided $t$-test, $p>0.05$), while DiffuASR has more than three times the across-seed standard deviation of DA-RSIR-FT ($0.10$ versus $0.03$). On Recall@10 in the same setting, DA-RSIR has the highest mean, while DiffuASR and DA-RSIR-FT perform nearly identically.

\subsection{Recursive Performance and Disagreement-Based Selection}
\label{sec:analysis}

\begin{figure*}[t]
\centering

\begin{minipage}[t]{0.42\textwidth}
\vspace{0pt}
\centering
\captionof{table}{Disagreement-regime comparison on Amazon Beauty with SASRec.
All policies use the same verified bank and select three candidates per proposal
group. Four-run means, multiplied by $100$; best in bold.}
\label{tab:regime-ablation}
\vspace{3pt}
\scriptsize
\setlength{\tabcolsep}{4pt}
\begin{tabular}{@{}lcc@{}}
\toprule
Policy & N@10 & R@10 \\
\midrule
Random-3 & 5.12 & 8.79 \\
Low-3    & 5.09 & 8.74 \\
Middle-3 & 5.14 & 8.81 \\
High-3   & 5.11 & 8.84 \\
\midrule
DA-RSIR  & \textbf{5.17} & \textbf{8.86} \\
\bottomrule
\end{tabular}
\end{minipage}
\hfill
\begin{minipage}[t]{0.56\textwidth}
\vspace{0pt}
\centering

\begin{minipage}[t]{0.48\linewidth}
\centering
\includegraphics[width=\linewidth]{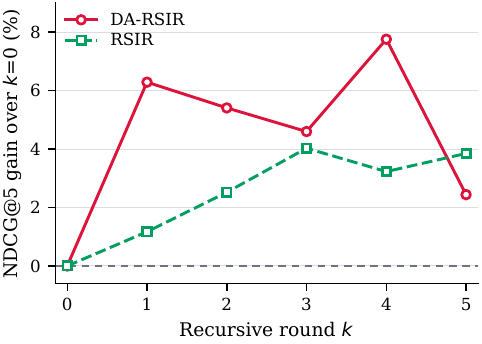}\\
{\small (a) NDCG@5}
\end{minipage}
\hfill
\begin{minipage}[t]{0.48\linewidth}
\centering
\includegraphics[width=\linewidth]{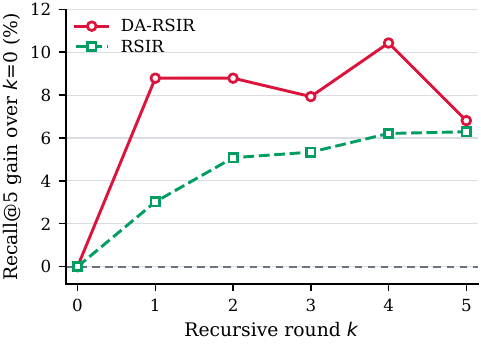}\\
{\small (b) Recall@5}
\end{minipage}

\captionof{figure}{Relative improvement over the initial model across five
recursive rounds for RSIR and DA-RSIR on Amazon Sports for SASRec model.}
\label{fig:recursive-rounds}
\end{minipage}

\end{figure*}

We examine whether the effect of acquisition persists across recursive rounds. Figure~\ref{fig:recursive-rounds} follows RSIR and DA-RSIR for five rounds on Amazon Sports using NDCG@5 and Recall@5 in a single-seed analysis. After one round, DA-RSIR improves over the initial model by $6.3\%$ in NDCG@5 and $8.8\%$ in Recall@5, exceeding the best gains reached by RSIR over five rounds. DA-RSIR remains above RSIR in all five post-initialization rounds for Recall@5 and four of five for NDCG@5. Performance peaks at round $4$ and declines at round $5$, a pattern consistent with the fidelity-leakage noise floor analyzed by \citet{zhang2026rsir}.

To separate disagreement-based selection from limiting the acquired increment, Table~\ref{tab:regime-ablation} compares policies on the same verified bank, with each policy acquiring exactly three candidates per proposal group. Random-3 samples uniformly from the full group, while Low-3, Middle-3, and High-3 select from the corresponding parts of the disagreement ranking; High-3 is a top-score-only strategy. Candidate generation, fidelity verification, and successor training remain unchanged. Random-3 outperforms both Base and retain-all RSIR, showing that limiting the acquired increment contributes to the improvement. Among the single-regime policies, Middle-3 has the highest NDCG@10 and High-3 the highest Recall@10, so no single disagreement regime dominates both metrics. DA-RSIR, which samples across all three regimes, attains the highest mean on both metrics, although differences among the fixed-count policies are modest relative to across-seed variability.

\section{Conclusion}

We formulate post-verification acquisition as a distinct control problem in recursive self-improving recommendation, separating which candidates pass verification from which verified candidates are used to train the next model. We introduce DA-RSIR, which preserves RSIR's generation and verification, limits each source sequence's contribution, and uses posterior disagreement to sample candidates across low-, middle-, and high-disagreement ranges. Across four datasets and three recommender models, DA-RSIR outperforms retain-all RSIR in all $24$ retraining comparisons.

Our experiments show that limiting the acquired increment contributes to the improvement, while no single disagreement regime dominates both metrics; DA-RSIR attains the highest mean on both. Within the five-round evaluation, one DA-RSIR update exceeds the best gain reached by retain-all RSIR. Together, these findings show that candidate selection after verification affects both immediate performance and later recursive updates. Broader impacts are discussed in Appendix~\ref{sec:broader_impacts}.

\section{Limitation and Future Work}
\label{sec:limitations}

DA-RSIR is evaluated on ID-based sequential recommender systems and does not consider text-based recommendation. We establish post-verification acquisition on four sparse, short-history benchmarks with three attention-based models; extending the study to long-history, session-based, and industrial-scale corpora with million-scale catalogs, as well as to non-attention and language-model-based recommenders, is natural future work. Our evaluation uses offline ranking metrics, so online evaluation remains to be studied. DA-RSIR samples across disagreement ranges using the same acquisition rule at every recursive round; future work can study when particular ranges should receive greater weight and whether acquisition should adapt across rounds.

\bibliographystyle{plainnat}
\bibliography{main}

@inproceedings{alemohammad2024self,
  title     = {Self-Consuming Generative Models Go {MAD}},
  author    = {Alemohammad, Sina and Casco-Rodriguez, Josue and Luzi, Lorenzo and
               Humayun, Ahmed Imtiaz and Babaei, Hossein and LeJeune, Daniel and
               Siahkoohi, Ali and Baraniuk, Richard G.},
  booktitle = {International Conference on Learning Representations (ICLR)},
  pages     = {53581--53608},
  year      = {2024}
}

@inproceedings{arazo2020pseudo,
  title        = {Pseudo-Labeling and Confirmation Bias in Deep Semi-Supervised Learning},
  author       = {Arazo, Eric and Ortego, Diego and Albert, Paul and
                  O'Connor, Noel E. and McGuinness, Kevin},
  booktitle    = {International Joint Conference on Neural Networks (IJCNN)},
  year         = {2020},
  organization = {IEEE}
}

@article{dang2024survey,
  title   = {Data Augmentation for Sequential Recommendation: A Survey},
  author  = {Dang, Yizhou and Yang, Enneng and Liu, Yuting and Guo, Guibing and
             Jiang, Linying and Zhao, Jianzhe and Wang, Xingwei},
  journal = {arXiv preprint arXiv:2409.13545},
  year    = {2024}
}

@inproceedings{du2023frequency,
  title     = {Frequency Enhanced Hybrid Attention Network for Sequential
               Recommendation},
  author    = {Du, Xinyu and Yuan, Huanhuan and Zhao, Pengpeng and Qu, Jianfeng and
               Zhuang, Fuzhen and Liu, Guanfeng and Sheng, Victor S.},
  booktitle = {Proceedings of the 46th International ACM SIGIR Conference on Research
               and Development in Information Retrieval (SIGIR)},
  pages     = {78--88},
  year      = {2023},
  doi       = {10.1145/3539618.3591689}
}

@inproceedings{gal2016dropout,
  title     = {Dropout as a {B}ayesian Approximation: Representing Model Uncertainty
               in Deep Learning},
  author    = {Gal, Yarin and Ghahramani, Zoubin},
  booktitle = {International Conference on Machine Learning (ICML)},
  year      = {2016}
}

@inproceedings{gal2017deep,
  title     = {Deep {B}ayesian Active Learning with Image Data},
  author    = {Gal, Yarin and Islam, Riashat and Ghahramani, Zoubin},
  booktitle = {International Conference on Machine Learning (ICML)},
  year      = {2017}
}

@inproceedings{gerstgrasser2024model,
  title     = {Is Model Collapse Inevitable? Breaking the Curse of Recursion by
               Accumulating Real and Synthetic Data},
  author    = {Gerstgrasser, Matthias and Schaeffer, Rylan and Dey, Apratim and
               Rafailov, Rafael and Korbak, Tomasz and Sleight, Henry and
               Agrawal, Rajashree and Hughes, John and Pai, Dhruv Bhandarkar and
               Gromov, Andrey and Roberts, Dan and Yang, Diyi and Donoho, David L.
               and Koyejo, Sanmi},
  booktitle = {First Conference on Language Modeling (COLM)},
  year      = {2024}
}

@inproceedings{ghorbani2019datashapley,
  title     = {Data Shapley: Equitable Valuation of Data for Machine Learning},
  author    = {Ghorbani, Amirata and Zou, James},
  booktitle = {Proceedings of the 36th International Conference on Machine
               Learning (ICML)},
  series    = {Proceedings of Machine Learning Research},
  volume    = {97},
  pages     = {2242--2251},
  year      = {2019},
  publisher = {PMLR}
}

@article{gulcehre2023rest,
  title   = {Reinforced Self-Training ({ReST}) for Language Modeling},
  author  = {Gulcehre, Caglar and Paine, Tom Le and Srinivasan, Srivatsan and
             Konyushkova, Ksenia and Weerts, Lotte and Sharma, Abhishek and
             Siddhant, Aditya and Ahern, Alex and Wang, Miaosen and Gu, Chenjie and
             Macherey, Wolfgang and Doucet, Arnaud and Firat, Orhan and
             de Freitas, Nando},
  journal = {arXiv preprint arXiv:2308.08998},
  year    = {2023}
}

@inproceedings{hidasi2016session,
  title     = {Session-Based Recommendations with Recurrent Neural Networks},
  author    = {Hidasi, Bal{\'a}zs and Karatzoglou, Alexandros and Baltrunas, Linas
               and Tikk, Domonkos},
  booktitle = {International Conference on Learning Representations (ICLR)},
  year      = {2016}
}

@article{houlsby2011bald,
  title   = {Bayesian Active Learning for Classification and Preference Learning},
  author  = {Houlsby, Neil and Husz{\'a}r, Ferenc and Ghahramani, Zoubin and
             Lengyel, M{\'a}t{\'e}},
  journal = {arXiv preprint arXiv:1112.5745},
  year    = {2011}
}

@article{idrissi2020systematic,
  title   = {A Systematic Literature Review of Sparsity Issues in Recommender Systems},
  author  = {Idrissi, Nouhaila and Zellou, Ahmed},
  journal = {Social Network Analysis and Mining},
  volume  = {10},
  number  = {1},
  pages   = {15},
  year    = {2020}
}

@inproceedings{kang2018selfattentive,
  title     = {Self-Attentive Sequential Recommendation},
  author    = {Kang, Wang-Cheng and McAuley, Julian},
  booktitle = {IEEE International Conference on Data Mining (ICDM)},
  pages     = {197--206},
  year      = {2018},
  publisher = {IEEE},
  doi       = {10.1109/ICDM.2018.00035}
}

@inproceedings{kirsch2019batchbald,
  title     = {{BatchBALD}: Efficient and Diverse Batch Acquisition for Deep
               {B}ayesian Active Learning},
  author    = {Kirsch, Andreas and van Amersfoort, Joost and Gal, Yarin},
  booktitle = {Advances in Neural Information Processing Systems (NeurIPS)},
  year      = {2019}
}

@inproceedings{koh2017understanding,
  title     = {Understanding Black-Box Predictions via Influence Functions},
  author    = {Koh, Pang Wei and Liang, Percy},
  booktitle = {Proceedings of the 34th International Conference on Machine
               Learning (ICML)},
  series    = {Proceedings of Machine Learning Research},
  volume    = {70},
  pages     = {1885--1894},
  year      = {2017},
  publisher = {PMLR}
}

@inproceedings{liu2021augmenting,
  title     = {Augmenting Sequential Recommendation with Pseudo-Prior Items via
               Reversely Pre-Training Transformer},
  author    = {Liu, Zhiwei and Fan, Ziwei and Wang, Yu and Yu, Philip S.},
  booktitle = {Proceedings of the 44th International ACM SIGIR Conference on Research
               and Development in Information Retrieval (SIGIR)},
  pages     = {1608--1612},
  year      = {2021},
  doi       = {10.1145/3404835.3463036}
}

@inproceedings{liu2023diffusion,
  title     = {Diffusion Augmentation for Sequential Recommendation},
  author    = {Liu, Qidong and Yan, Fan and Zhao, Xiangyu and Du, Zhaocheng and
               Guo, Huifeng and Tang, Ruiming and Tian, Feng},
  booktitle = {Proceedings of the 32nd ACM International Conference on Information
               and Knowledge Management (CIKM)},
  pages     = {1576--1586},
  year      = {2023},
  doi       = {10.1145/3583780.3615134}
}

@inproceedings{mcauley2015image,
  title     = {Image-Based Recommendations on Styles and Substitutes},
  author    = {McAuley, Julian and Targett, Christopher and Shi, Qinfeng and
               van den Hengel, Anton},
  booktitle = {Proceedings of the 38th International ACM SIGIR Conference on Research
               and Development in Information Retrieval (SIGIR)},
  pages     = {43--52},
  year      = {2015},
  doi       = {10.1145/2766462.2767755}
}

@inproceedings{sener2018active,
  title     = {Active Learning for Convolutional Neural Networks: A Core-Set Approach},
  author    = {Sener, Ozan and Savarese, Silvio},
  booktitle = {International Conference on Learning Representations (ICLR)},
  year      = {2018}
}

@article{shumailov2024collapse,
  title   = {{AI} Models Collapse When Trained on Recursively Generated Data},
  author  = {Shumailov, Ilia and Shumaylov, Zakhar and Zhao, Yiren and
             Papernot, Nicolas and Anderson, Ross and Gal, Yarin},
  journal = {Nature},
  volume  = {631},
  number  = {8022},
  pages   = {755--759},
  year    = {2024},
  doi     = {10.1038/s41586-024-07566-y}
}

@inproceedings{sohn2020fixmatch,
  title     = {{FixMatch}: Simplifying Semi-Supervised Learning with Consistency and
               Confidence},
  author    = {Sohn, Kihyuk and Berthelot, David and Carlini, Nicholas and
               Zhang, Zizhao and Zhang, Han and Raffel, Colin A. and
               Cubuk, Ekin Dogus and Kurakin, Alexey and Li, Chun-Liang},
  booktitle = {Advances in Neural Information Processing Systems (NeurIPS)},
  volume    = {33},
  pages     = {596--608},
  year      = {2020}
}

@inproceedings{sorscher2022beyond,
  title     = {Beyond Neural Scaling Laws: Beating Power Law Scaling via Data Pruning},
  author    = {Sorscher, Ben and Geirhos, Robert and Shekhar, Shashank and
               Ganguli, Surya and Morcos, Ari},
  booktitle = {Advances in Neural Information Processing Systems (NeurIPS)},
  volume    = {35},
  pages     = {19523--19536},
  year      = {2022}
}

@inproceedings{sun2019bert4rec,
  title     = {{BERT4Rec}: Sequential Recommendation with Bidirectional Encoder
               Representations from Transformer},
  author    = {Sun, Fei and Liu, Jun and Wu, Jian and Pei, Changhua and Lin, Xiao and
               Ou, Wenwu and Jiang, Peng},
  booktitle = {Proceedings of the 28th ACM International Conference on Information
               and Knowledge Management (CIKM)},
  pages     = {1441--1450},
  year      = {2019},
  doi       = {10.1145/3357384.3357895}
}

@inproceedings{tang2018personalized,
  title     = {Personalized Top-N Sequential Recommendation via Convolutional Sequence
               Embedding},
  author    = {Tang, Jiaxi and Wang, Ke},
  booktitle = {Proceedings of the 11th ACM International Conference on Web Search and
               Data Mining (WSDM)},
  pages     = {565--573},
  year      = {2018},
  doi       = {10.1145/3159652.3159656}
}

@inproceedings{wang2023selfinstruct,
  title     = {Self-Instruct: Aligning Language Models with Self-Generated
               Instructions},
  author    = {Wang, Yizhong and Kordi, Yeganeh and Mishra, Swaroop and Liu, Alisa and
               Smith, Noah A. and Khashabi, Daniel and Hajishirzi, Hannaneh},
  booktitle = {Proceedings of the 61st Annual Meeting of the Association for
               Computational Linguistics (ACL)},
  pages     = {13484--13508},
  year      = {2023},
  publisher = {Association for Computational Linguistics},
  doi       = {10.18653/v1/2023.acl-long.754}
}

@inproceedings{xia2024less,
  title     = {{LESS}: Selecting Influential Data for Targeted Instruction Tuning},
  author    = {Xia, Mengzhou and Malladi, Sadhika and Gururangan, Suchin and
               Arora, Sanjeev and Chen, Danqi},
  booktitle = {Proceedings of the 41st International Conference on Machine
               Learning (ICML)},
  series    = {Proceedings of Machine Learning Research},
  volume    = {235},
  pages     = {54104--54132},
  year      = {2024},
  publisher = {PMLR}
}

@inproceedings{xie2022contrastive,
  title     = {Contrastive Learning for Sequential Recommendation},
  author    = {Xie, Xu and Sun, Fei and Liu, Zhaoyang and Wu, Shiwen and
               Gao, Jinyang and Zhang, Jiandong and Ding, Bolin and Cui, Bin},
  booktitle = {IEEE International Conference on Data Engineering (ICDE)},
  pages     = {1259--1273},
  year      = {2022},
  publisher = {IEEE},
  doi       = {10.1109/ICDE53745.2022.00099}
}

@misc{yelp_open_dataset,
  title        = {Yelp Open Dataset},
  author       = {{Yelp Inc.}},
  howpublished = {\url{https://business.yelp.com/data/resources/open-dataset/}},
  year         = {2024}
}

@inproceedings{yin2024dataset,
  title     = {Dataset Regeneration for Sequential Recommendation},
  author    = {Yin, Mingjia and Wang, Hao and Guo, Wei and Liu, Yong and
               Zhang, Suojuan and Zhao, Sirui and Lian, Defu and Chen, Enhong},
  booktitle = {Proceedings of the 30th ACM SIGKDD Conference on Knowledge Discovery
               and Data Mining (KDD)},
  pages     = {3954--3965},
  year      = {2024},
  doi       = {10.1145/3637528.3671841}
}

@inproceedings{zelikman2022star,
  title     = {{STaR}: Bootstrapping Reasoning with Reasoning},
  author    = {Zelikman, Eric and Wu, Yuhuai and Mu, Jesse and Goodman, Noah D.},
  booktitle = {Advances in Neural Information Processing Systems (NeurIPS)},
  volume    = {35},
  pages     = {15476--15488},
  year      = {2022}
}

@inproceedings{zhai2024actions,
  title     = {Actions Speak Louder than Words: Trillion-Parameter Sequential
               Transducers for Generative Recommendations},
  author    = {Zhai, Jiaqi and Liao, Lucy and Liu, Xing and Wang, Yueming and Li, Rui
               and Cao, Xuan and Gao, Leon and Gong, Zhaojie and Gu, Fangda and
               He, Jiayuan and Lu, Yinghai and Shi, Yu},
  booktitle = {Proceedings of the 41st International Conference on Machine
               Learning (ICML)},
  series    = {Proceedings of Machine Learning Research},
  volume    = {235},
  pages     = {58484--58509},
  year      = {2024},
  publisher = {PMLR}
}

@inproceedings{zhang2026rsir,
  title     = {Can Recommender Systems Teach Themselves? A Recursive Self-Improving
               Framework with Fidelity Control},
  author    = {Zhang, Luankang and Wang, Hao and Liu, Zhongzhou and Yin, Mingjia and
               Huang, Yonghao and Li, Jiaqi and Guo, Wei and Liu, Yong and
               Guo, Huifeng and Lian, Defu and Chen, Enhong},
  booktitle = {Proceedings of the 43rd International Conference on Machine
               Learning (ICML)},
  series    = {Proceedings of Machine Learning Research},
  volume    = {306},
  year      = {2026}
}

@inproceedings{zhou2020s3rec,
  title     = {S3-Rec: Self-Supervised Learning for Sequential Recommendation with
               Mutual Information Maximization},
  author    = {Zhou, Kun and Wang, Hui and Zhao, Wayne Xin and Zhu, Yutao and
               Wang, Sirui and Zhang, Fuzheng and Wang, Zhongyuan and Wen, Ji-Rong},
  booktitle = {Proceedings of the 29th ACM International Conference on Information
               and Knowledge Management (CIKM)},
  pages     = {1893--1902},
  year      = {2020},
  doi       = {10.1145/3340531.3411954}
}

@inproceedings{zhou2024contrastive,
  title     = {Is Contrastive Learning Necessary? A Study of Data Augmentation vs
               Contrastive Learning in Sequential Recommendation},
  author    = {Zhou, Peilin and Huang, You-Liang and Xie, Yueqi and Gao, Jingqi and
               Wang, Shoujin and Kim, Jae Boum and Kim, Sunghun},
  booktitle = {Proceedings of the ACM Web Conference 2024 (WWW)},
  pages     = {3854--3863},
  year      = {2024},
  doi       = {10.1145/3589334.3645661}
}

\clearpage
\appendix

\section{DA-RSIR Acquisition Algorithm}
\label{app:algorithm}

Algorithm~\ref{alg:darsir} summarizes the overall DA-RSIR acquisition procedure. It scores verified candidates, orders them within each source sequence and proposal group, and selects representatives from the disagreement regimes.

\begin{algorithm}[H]
\caption{DA-RSIR acquisition at round $k$}
\label{alg:darsir}
\begin{algorithmic}[1]
\Require Fidelity-valid bank $\mathcal{C}^{\mathrm{fid}}_{k+1}$ grouped by source sequence and proposal group, current model $\theta_k$, MC samples $S$, experiment seed
\Ensure Synthetic increment $\mathcal{A}_{k+1}$

\State $\mathcal{A}\gets\varnothing$

\ForAll{$z\in\mathcal{C}^{\mathrm{fid}}_{k+1}$}
    \State compute $U_{\mathrm{mean}}(z)$, maximum target-specific disagreement, and mean posterior-predictive log-likelihood
\EndFor

\ForAll{nonempty proposal groups $G$}
    \State sort $G$ by lower $U_{\mathrm{mean}}$, lower maximum disagreement, higher posterior likelihood, attempt index, and complete sequence
    \State $n\gets|G|$
    \If{$n\leq3$}
        \State $\mathcal{R}\gets G$
    \Else
        \State $h\gets\max\{1,\lfloor n/4\rfloor\}$
        \State form low, middle, and high regimes from the first $h$, remaining, and last $h$ candidates
        \State $\mathcal{R}\gets$ one uniform sample from each regime using the keyed seed
    \EndIf
    \State $\mathcal{A}\gets\mathcal{A}\cup\mathcal{R}$
\EndFor

\State remove exact duplicate sequences from $\mathcal{A}$
\State \textbf{return} $\mathcal{A}_{k+1}\gets\mathcal{A}$
\end{algorithmic}
\end{algorithm}

\section{Additional Experimental Details}
\label{app:experimental_details}

\paragraph{\bf Datasets and preprocessing.}
We evaluate DA-RSIR on four public sequential-recommendation benchmarks: Amazon Toys, Amazon Beauty, and Amazon Sports from the Amazon Review Data \citep{mcauley2015image}, and Yelp from the Yelp Open Dataset \citep{yelp_open_dataset}.
Following RSIR \citep{zhang2026rsir}, we apply 5-core preprocessing so that every retained user and item participates in at least five interactions. All four benchmarks are highly sparse, with average sequence lengths between $5.3$ and $7.1$ items, which is the regime in which synthetic augmentation is typically applied. We order each user's remaining interactions chronologically and use the same preprocessed datasets and data splits for all compared methods.
Dataset statistics after preprocessing are reported in Section~\ref{app:dataset_statistics}.

\paragraph{\bf Evaluation protocol and metrics.}
We follow the leave-one-out protocol, reserving the last interaction of each user sequence for testing, the second-to-last interaction for validation, and all preceding interactions for training.
We report Recall@10 and NDCG@10, computed by ranking the held-out item against the full item catalog.
All reported values are multiplied by $100$ for readability.
We report the mean and sample standard deviation over four random seeds.

\paragraph{\bf Hyperparameters and implementation details.}
We follow the tuning protocol of \citet{zhang2026rsir} for the inherited generation and verification hyperparameters, selecting values by validation NDCG@10 over the grids reported in that work. These are the fidelity rank threshold $\tau$, the local-candidate probability $p_{\mathrm{loc}}$, the top-$\kappa$ sampling cutoff, and the minimum retained sequence length. The threshold $\tau$ determines how highly at least one remaining item from the source sequence must be ranked for a proposed extension to pass fidelity verification; $p_{\mathrm{loc}}$ controls whether the candidate set is sampled from the source sequence or the global item catalog; and $\kappa$ determines how many highest-scoring candidate items remain eligible for sampling. The selected configuration is shared by RSIR and DA-RSIR, so each matched pair differs only in the post-verification acquisition rule.

The DA-RSIR-specific hyperparameters are the number of proposal groups $B$, generation attempts per group $M$, and Monte Carlo dropout samples $S$, with $BM$ total generation attempts per source sequence. We evaluate $B\in\{1,2,3,4,5,6\}$, $M\in\{10,15,20,25,30\}$, and $S\in\{5,10,20,30,40,50\}$ using validation performance. The final configuration uses $B=2$, $M=10$, and $S=10$, giving $20$ generation attempts per source sequence. The case $B=1$ corresponds to source-sequence-conditioned acquisition without proposal grouping. For each model, all successor models use the same architecture, embedding dimension, optimization procedure, negative-sampling strategy, maximum sequence length, and early-stopping criterion. We optimize the next-item prediction objective using sampled softmax with $256$ negative items per training position and, unless otherwise stated, evaluate performance after one recursive self-improvement round.

All experiments were conducted on a single compute node with eight NVIDIA H100 GPUs. Aggregated across the four datasets, generation, retraining, and fine-tuning required approximately $25$, $35$, and $34$ minutes for SASRec; $26$, $312$, and $85$ minutes for CL4SRec; and $32$, $117$, and $88$ minutes for FEARec, corresponding to approximately $1.6$, $7.0$, and $4.0$ GPU-hours, respectively. In proposal-matched controls, all acquisition policies share the same model checkpoint, source sequences, generated proposals, generation seeds, decoding configuration, fidelity decisions, minimum-length filtering, sequence deduplication, and successor-training procedure; only the post-verification acquisition rule differs.

\section{Additional Result Analyses}
\label{app:sensitivity}

This section provides additional analyses of two DA-RSIR design choices: the number of proposal groups $B$ and Monte Carlo dropout samples $S$. We examine how each choice affects performance and assess the sensitivity of DA-RSIR to these settings.

\subsection{Sensitivity to Proposal Grouping}
\label{app:proposal_grouping}

\begin{figure}[t]
\centering

\begin{minipage}[t]{0.48\linewidth}
\centering
\includegraphics[width=\linewidth]{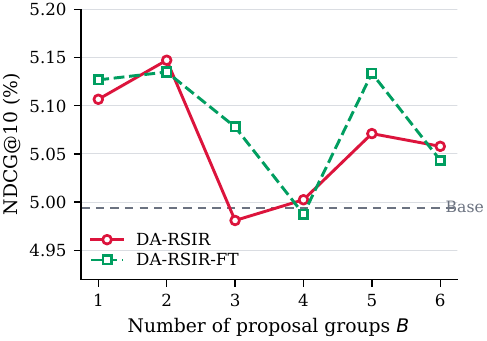}\\
{\small (a) NDCG@10}
\end{minipage}
\hfill
\begin{minipage}[t]{0.48\linewidth}
\centering
\includegraphics[width=\linewidth]{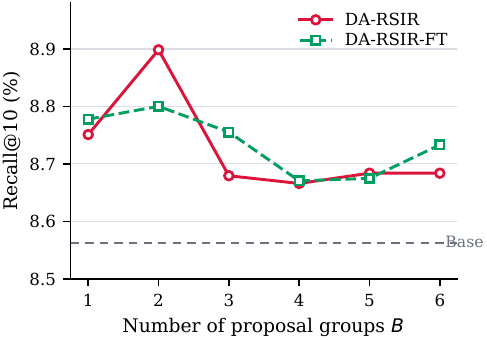}\\
{\small (b) Recall@10}
\end{minipage}

\caption{Effect of the number of proposal groups $B$ on Amazon Beauty with SASRec while holding the acquired count fixed at $12{,}897$ sequences from a shared proposal bank.}
\label{fig:beauty-block-ablation}
\end{figure}

Figure~\ref{fig:beauty-block-ablation} varies the number of proposal groups $B$ while holding the acquired count fixed at $12{,}897$ sequences from a shared proposal bank, isolating grouping from acquisition volume. Performance varies non-monotonically with $B$. Under retraining, $B=2$ attains the highest NDCG@10 and Recall@10. Under fine-tuning, $B=2$ attains the highest Recall@10, while $B=5$ yields a marginally higher NDCG@10. Every tested $B$ improves Recall@10 over Base, whereas retraining with $B=3$ and $B=4$ returns approximately to the Base NDCG@10 level. The competitive performance of $B=1$ suggests that source-sequence-conditioned ranking captures much of the benefit, while additional proposal grouping provides a modest and setting-dependent gain.

\subsection{Sensitivity to Monte Carlo Sample Count}
\label{app:mc_samples}

Table~\ref{tab:s-ablation} varies the number of Monte Carlo dropout samples $S$. The default $S=10$ attains the highest mean in all four metric and training-procedure combinations, while increasing $S$ beyond $10$ provides no further improvement. Performance remains stable across the tested range. Although $S=5$ remains competitive with half as many stochastic forward passes, we retain $S=10$ because it achieves the highest mean consistently across both metrics and training procedures.

\begin{table}[t]
\centering
\caption{Effect of the number of Monte Carlo dropout samples $S$ on Amazon Beauty with SASRec. Means over four seeds, $\times100$; best value in each column in bold.}
\label{tab:s-ablation}
\setlength{\tabcolsep}{3pt}
\begin{tabular}{@{}lcccc@{}}
\toprule
& \multicolumn{2}{c}{\textbf{DA-RSIR}} 
& \multicolumn{2}{c}{\textbf{DA-RSIR-FT}} \\
\cmidrule(lr){2-3} \cmidrule(lr){4-5}
$S$ & N@10 & R@10 & N@10 & R@10 \\
\midrule
5  & 5.06 & 8.70 & 5.08 & 8.74 \\
10 & \textbf{5.17} & \textbf{8.86} & \textbf{5.17} & \textbf{8.81} \\
20 & 5.09 & 8.72 & 5.07 & 8.74 \\
30 & 5.10 & 8.74 & 5.13 & 8.77 \\
40 & 5.03 & 8.77 & 5.08 & 8.75 \\
50 & 5.10 & 8.75 & 5.08 & 8.73 \\
\bottomrule
\end{tabular}
\end{table}

\section{Dataset and Acquired-Corpus Statistics}
\label{app:dataset_statistics}

Table~\ref{tab:original-dataset-statistics} reports the statistics of the four preprocessed datasets, while Table~\ref{tab:darsir-corpus-growth} reports the first recursive round corpus expansion produced by DA-RSIR. Across datasets and models, DA-RSIR increases the sequence count by $44.1\%$--$81.8\%$ and the interaction count by $30.2\%$--$63.3\%$. Because the user and item sets remain fixed, the method adds training trajectories within the existing recommendation domain rather than expanding the user or item sets. The different corpus sizes produced by the three models on the same dataset further show that verified candidate yield depends on the current model.

\begin{table}[t]
\centering
\caption{Statistics of the 5-core datasets after preprocessing. Each user
contributes one original interaction sequence. Avg. length denotes the mean
number of interactions per user.}
\label{tab:original-dataset-statistics}
\setlength{\tabcolsep}{3pt}
\begin{tabular}{lrrrrr}
\toprule
Dataset
& Users
& Items
& Interactions
& Avg. length
& Sparsity \\
\midrule
Amazon Toys
& 19,412
& 11,876
& 106,254
& 5.47
& 0.999539 \\

Amazon Beauty
& 22,363
& 12,066
& 127,598
& 5.71
& 0.999527 \\

Amazon Sports
& 35,598
& 18,281
& 187,694
& 5.27
& 0.999712 \\

Yelp
& 30,431
& 20,014
& 216,733
& 7.12
& 0.999644 \\
\bottomrule
\end{tabular}
\end{table}

\begin{table*}[t]
\centering
\caption{First-round DA-RSIR corpus expansion with
$B=2$. Added counts and percentages are measured relative to the original
training corpus. Interaction count denotes the total number of item positions
across all training sequences. The user and item sets remain unchanged.}
\label{tab:darsir-corpus-growth}
\vspace{3pt}
\small
\setlength{\tabcolsep}{4pt}
\begin{tabular}{llrrrr}
\toprule
Dataset
& Model
& Added sequences
& Total sequences
& Added interactions
& Total interactions \\
\midrule

\multirow{3}{*}{Amazon Toys}
& SASRec
& 13,364 {\scriptsize($+68.8\%$)}
& 32,776
& 51,047 {\scriptsize($+48.0\%$)}
& 157,301 \\

& CL4SRec
& 10,648 {\scriptsize($+54.9\%$)}
& 30,060
& 39,236 {\scriptsize($+36.9\%$)}
& 145,490 \\

& FEARec
& 13,207 {\scriptsize($+68.0\%$)}
& 32,619
& 47,110 {\scriptsize($+44.3\%$)}
& 153,364 \\

\midrule

\multirow{3}{*}{Amazon Beauty}
& SASRec
& 16,328 {\scriptsize($+73.0\%$)}
& 38,691
& 80,737 {\scriptsize($+63.3\%$)}
& 208,335 \\

& CL4SRec
& 18,290 {\scriptsize($+81.8\%$)}
& 40,653
& 75,668 {\scriptsize($+59.3\%$)}
& 203,266 \\

& FEARec
& 18,032 {\scriptsize($+80.6\%$)}
& 40,395
& 77,138 {\scriptsize($+60.5\%$)}
& 204,736 \\

\midrule

\multirow{3}{*}{Amazon Sports}
& SASRec
& 15,697 {\scriptsize($+44.1\%$)}
& 51,295
& 56,619 {\scriptsize($+30.2\%$)}
& 244,313 \\

& CL4SRec
& 22,847 {\scriptsize($+64.2\%$)}
& 58,445
& 73,879 {\scriptsize($+39.4\%$)}
& 261,573 \\

& FEARec
& 19,263 {\scriptsize($+54.1\%$)}
& 54,861
& 62,288 {\scriptsize($+33.2\%$)}
& 249,982 \\

\midrule

\multirow{3}{*}{Yelp}
& SASRec
& 18,165 {\scriptsize($+59.7\%$)}
& 48,596
& 70,250 {\scriptsize($+32.4\%$)}
& 286,983 \\

& CL4SRec
& 17,447 {\scriptsize($+57.3\%$)}
& 47,878
& 65,970 {\scriptsize($+30.4\%$)}
& 282,703 \\

& FEARec
& 23,998 {\scriptsize($+78.9\%$)}
& 54,429
& 88,054 {\scriptsize($+40.6\%$)}
& 304,787 \\

\bottomrule
\end{tabular}
\end{table*}

\end{document}